# Performance evaluation of an integrated photonic convolutional neural network based on delay buffering and wavelength division multiplexing

SHAOFU XU,[1] JING WANG,[1] AND WEIWEN ZOU[1,*]

[1]*State Key Laboratory of Advanced Optical Communication Systems and Networks, Intelligent Microwave Lightwave Integration Innovation Center (iMLic), Department of Electronic Engineering, Shanghai Jiao Tong University, 800 Dongchuan Road, Shanghai 200240, China*
*\*wzou@sjtu.edu.cn*

**Abstract:** Photonic technologies have shown a promising way to build high-speed and high-energy-efficiency neural network accelerators. In previously presented photonic neural networks, architectures are mainly designed for fully-connected layers. When convolutional layers are executed in such neural networks, the large-scale electrooptic modulation array heavily increases the energy dissipation on chip. To increase the energy efficiency, here we show an integrated photonic architecture specifically for convolutional layer calculations. Optical delay lines replace electronics to execute data manipulations on optical chip, reducing the scale of electro-optic modulation array. Consequently, the energy dissipation of these parts is mitigated. Powered by wavelength division multiplexing, the footprint of delay lines is significantly reduced compared with previous art, thus being practical to fabricate. We evaluate the potential performance of the proposed architecture with respect to component flaws in practical fabrications. According to the results, with well-controlled system insertion loss, energy efficiency of the proposed architecture would surpass previously presented works and the state-of-art electronic processors. We anticipate the proposed architecture is beneficial for future fast and energy-efficient convolutional neural network accelerators.

## 1. Introduction

In recent years, powered by dramatic development of computer science and the explosive data amount, deep learning stands out among artificial intelligence technologies across a broad range of applications, including computer vision, natural language processing, games, and scientific researches [1-3]. As one of the most widely deployed deep learning models, convolutional neural networks (CNNs) are especially effective to process regularized data inputs. With multiple convolutional layers, input data from audios, images or videos can be extracted to high-level features which improves the performance of various tasks [4-6]. Therefore, in modern deep learning technologies, a majority of tasks involving regularized data inputs adopts convolutional layers to extract features.

Together with the performance improvement by large-scale neural networks, the computation burden and power consumption are increasing conspicuously. To solve the problem of energy consumption while maintaining high computing speed, scientists of integrated circuits are making efforts to develop higher-energy-efficiency deep learning accelerators. As an epitome, Application-specific integrated circuits (ASICs) like ISSAC [7], DianNao [8], and tensor processing units (TPUs) [9] have achieved unprecedented speed and energy efficiency. However, the integrated electronic circuits always face an energy bottleneck of on-chip data registering and manipulations. As a result, the best performance limited by 1 pico-Joule per multiply and accumulation (pJ/MAC). Recently, photonic neural network accelerators are proposed, heralding a novel way to break through the electronic energy bottleneck. Taking advantage of integrated or free-space optical components, such as Mach-Zehnder interference array [10], micro-ring resonator array [11], space-light modulators [12-14], and 3-D printed diffraction plates [15], photonic hardwares can calculate the linear part (i.e. vector-matrix multiplication) of neural networks with ultra-low power consumption. Note that the linear part of neural networks takes the majority of power consumption in traditional computers. Therefore, photonic hardware is promising to realize the ultrahigh-energy-efficiency deep learning accelerators. Nevertheless, most of current photonic neural network architectures are focusing on fully-connected layers in neural networks. When they are adopted to conduct convolutional layers, their energy efficiency can be challenged. A convolutional layer can be regarded as a generalized matrix multiplication (GeMM) [16]. Before the matrix multiplications, the input data should be firstly manipulated to match the GeMM format. In current photonic architectures, the data manipulations are assisted with digital electronic circuits, which introduce extra latency. Besides, the large number of high-speed input electro-optic modulators causes substantial power consumption. Recently, photonic architectures especially for CNNs has been proposed [17, 18], there occurs the same energy efficiency limitations as described above.

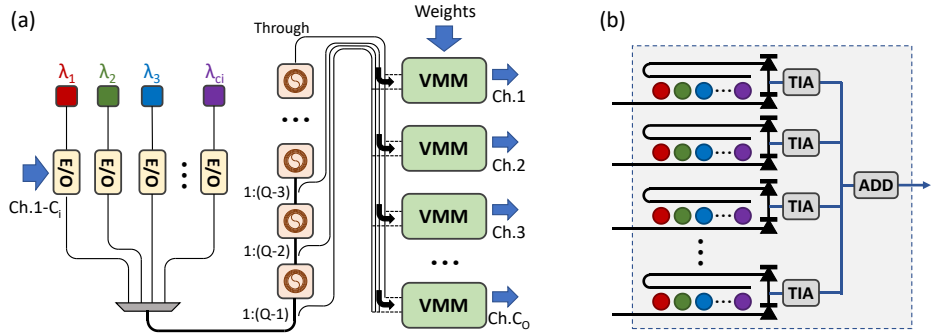

Fig. 1 Schematic of IPCNN. (a) The holistic schematic of the architecture. Input part contains $C_I$ laser sources with different wavelengths. High-speed electro-optic modulators (E/O) accept the serialized input data from $C_I$ input channels. With a WDM, wavelengths are combined into one waveguide and enter the delay line bank. After a delay line bank, the outputs are duplicated equally into $C_O$ VMM cores. Each VMM core finishes the computation of one output channel. (b) The structure of a VMM core. Each micro-ring resonator is tuned to control the coupling ratio between through port and drop port for a specific wavelength. And the intensities of different wavelengths are added up in the BPD and then amplified with transimpedance amplifiers (TIAs). Voltages are finally added up by a voltage adder (ADD).

In ref. [19], the authors presented a delay line method to execute data manipulations on optical chip. Following this idea, we present an energy-efficient integrated photonic architecture, IPCNN, especially for the calculation of convolutional layers in neural networks. We adopt optical delay lines as data buffer, replacing the electronic circuit to execute data patching and allocation. These data manipulations are processed with the speed of light and free of power. Due to the fact that data manipulations are executed in photonic hardwares, the proposed architecture requires less input electro-optic modulators, thus enhancing the energy-efficiency further. In contrast to [19], IPCNN adopts wavelength-division multiplexing (WDM) to combine multiple group of delay lines into one group. Therefore, it solves the problem of delay line footprint and shows a more practical way for chip fabrication. In this paper, we present the theoretical model of IPCNN to show its equivalence to GeMM-based neural networks. Then we evaluate the system performances including prediction accuracy, maximal integration scale, computing speed, and energy efficiency of IPCNN with respect to practical fabrication flaws like noise, imbalance, and insertion loss.

## 2. The IPCNN architecture

The IPCNN is shown in Fig. 1(a). The illustrated structure is equivalent to a convolutional layer with $C_I$ input channels and $C_O$ output channels, and the size of convolutional kernels is $Q$. For every input channel, a laser with a specific wavelength and a high-speed electro-optic (E/O) modulator is deployed. Totally $C_I$ wavelengths are deployed for the simultaneous data modulation of all input channels. Then, these wavelengths are combined into one waveguide with a wavelength-divided multiplexer (WDM) to reduce the usage of optical delay line. In the delay line bank, $(Q-1)$ delay lines are cascaded with a drop port between each. The physical delay of a dropped light is the accumulated traveling length before it is dropped. By designing the lengths of delay lines and the coupling ratio of the drop ports, we can obtain $Q$ intensity-equal copies of the input with different physical delays. In this way, the required the data manipulations before matrix multiplications are finished. Then, the matrix multiplication is implemented within the vector-matrix multiplication (VMM) cores. The outputs of the delay line bank are equally divided into $C_O$ copies, and each copy enters a VMM core to get the convolution result of an output channel. The structure of VMM core is depicted in Fig. 1(b) and the concept of this structure is presented in [11, 16]. Each VMM core has $Q$ input waveguides inside each travel $C_I$ wavelengths. In a VMM core, there are totally $C_I Q$ parameters represented by the micro-ring resonators (MRRs). A single micro-ring resonator acts as a tunable coupler for a specific wavelength, tuned to control the coupling ratio from through-port to drop-port: therefore, the intensity of the wavelength is weighted. Then, the intensities of $C_I$ weighted wavelengths are summed up and converted to electrical voltage in the balanced photo-detector (BPD) and transimpedance amplifier (TIA). All $Q$ voltages are summed up in the voltage adder to give the output of a VMM core. With a single VMM core, we can get the result of one output channel; with $C_O$ VMM cores, the entire convolutional layer is completed.

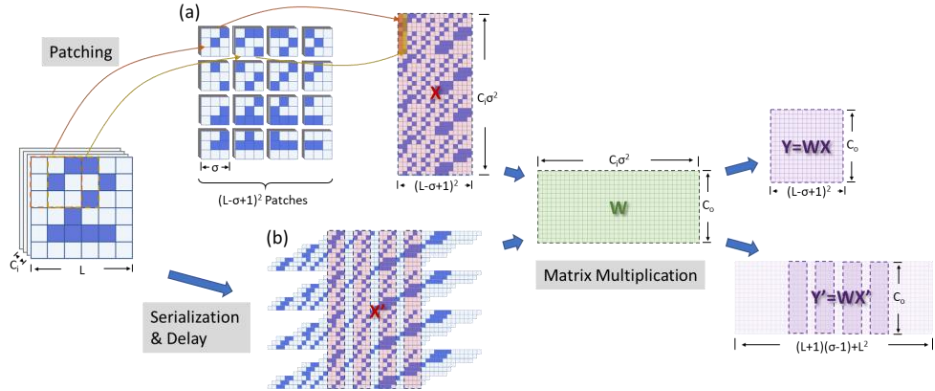

Fig. 2 The processes of GeMM for a convolutional layer. The convolutional layer shown here transforms $C_I$ input channels to $C_O$ output channels with 3×3-sized convolutional kernels. The image size is $L \times L$. **a,** The conventional way of GeMM in digital computers and previous photonic neural network architectures. Processes include patching, tiling and matrix multiplications. Patching and tiling are executed with electronic circuits in previous photonic neural networks. **b,** The IPCNN way to build the input matrix by serialization and delay. By imposing fixed delay amounts on input data, optical passive delay line bank can provide the input matrix $X'$. The valid part of input matrix $X'$ is equivalent to the conventional input matrix $X$ (shown in reddish part). After matrix multiplications, the valid part of the output matrix $Y'$ is equivalent to conventional output matrix $Y$ (shown in purple part).

In Fig. 2, we visualize the GeMM of a convolutional layer with image inputs to show its equivalence to delay buffering. Suppose the input data is a group of square-shaped images with width $L$ and channel number of $C_I$, and the convolutional layer is to convolute the $C_I$-channel input data to $C_O$-channel output data. Prior to matrix multiplications, the image data are preprocessed to form an input matrix. The first step is patching: dividing the input images into kernel-sized small patches with a fixed stride. In Fig. 2(a), we take the size of the convolutional kernels to be $Q=\sigma^2=3^2$, and the patching stride is set to 1. Therefore, a 6×6 image is divided to

16 patches. The patching process is repeated for every input channel. Then, each patch is flattened to form a piece of the input matrix X. Patches from the same input channel shall rank horizontally and different input channel are arranged vertically. Consequently, the size of matrix X is $(L-\sigma+1)^2 \times C_I\sigma^2$. In a single convolutional layer, the number of convolutional kernels is $C_IC_O$ so the aggregate number of parameters is $C_IC_O\sigma^2$. The convolutional kernels should also be transformed to form a weight matrix W with size $C_I\sigma^2 \times C_O$. With above-mentioned matrices X and W, the convolution result is represented with $Y=WX$ in matrix multiplication way. Each row of the result matrix Y can be transformed back to an output image channel. Therefore, a standard GeMM is finished.

In fact, there is another way to form the input matrix by delay buffering. The 2-dimensional (2-D) image convolution can be expressed by

$$y_{v,m,n} = \sum_{u=0}^{C_I}\sum_{i=0}^{\sigma-1}\sum_{j=0}^{\sigma-1}(w_{u,v,i,j} \cdot x_{u,m+i,n+j}), \tag{1}$$

where $y_{v,m,n}$ is the value of $v$-th output channel at the position of $(m, n)$, $w_{u,v,i,j}$ is the kernel value of $u$-th input channel to $v$-th output channel at position of $(i, j)$, $x_{u,m+i,n+j}$ is the image value of $u$-th input image at the position of $(m+i, n+j)$. By serializing input 2-D images and 2-D kernels row by row into 1-D sequences, we can use a single index to describe the position. Defining $q=i\sigma+j$, $s=mL+n$, we can rewrite Eq. (1) as follows.

$$w_{u,v,i,j} = \overline{w}_{u,v,i\sigma+j} = \overline{w}_{u,v,q} \tag{2}$$

$$x_{u,m+i,n+j} = \overline{x}_{u,(m+i)L+n+j} = \overline{x}_{u,s+iL+j} = \overline{x}_{u,s+\left\lfloor\frac{q}{\sigma}\right\rfloor L+\text{mod}(q,\sigma)} \tag{3}$$

$$y_{v,m,n} = \overline{y}_{v,mL+n} = \overline{y}_{v,s} = \sum_{u=0}^{C_I}\sum_{q=0}^{\sigma^2-1}(\overline{w}_{u,v,q} \cdot \overline{x}_{u,s+\left\lfloor\frac{q}{\sigma}\right\rfloor L+\text{mod}(q,\sigma)}) \tag{4}$$

All indices above are integers starting from 0. $\overline{w}$, $\overline{x}$, $\overline{y}$ are the 1-D serialized version of original 2-D kernels, input images, and output values, respectively. Equation (4) indicates 2-D image convolutions can be equivalently converted to 1-D sequence convolutions. As is shown in Fig. 2(b), we firstly serialize an input image into a row, and then duplicate it for Q copies, which are imposed with different fixed delay amounts. The delay amount of $q$-th copy is defined by Eq. (4)

$$D_q = \left\lfloor\frac{q}{\sigma}\right\rfloor L + \text{mod}(q,\sigma), q \in [0, \sigma^2 - 1]. \tag{5}$$

After every input channel is manipulated in the same way, we can get the delayed matrix X'. In X', we visualize its valid part by reddish zone, which is the same with X. In 2-D image convolutions, one position is valid if the position is inside the image, therefore, the valid part is determined by

$$\forall i, j \in [0, \sigma-1], m+i \in [0, L-1], n+j \in [0, L-1], \tag{6}$$

$$X'_{valid} = \{X'(:,s) \mid s = mL+n; m, n \in [0, L-\sigma]\}. \tag{7}$$

Note that the physical delay on the hardware is related to modulation rate of the input E/O, i.e. faster modulation rate requires shorter physical length of the delay line to get the required delay amount. If $f_m$ denotes the modulation rate, the physical delay is

$$D_q^{phy} = D_q / f_m. \tag{8}$$

Since the height of delayed matrix is also $C_i\sigma^2$, it can be multiplied with the weight matrix W. As a consequence, the rendered parts of the output matrix Y' is the same with the standard

GeMM output *Y*. The valid part (reddish part) of delayed matrix *X'* gives valid output (purple part in *Y'*).

## 3. Performance evaluations

### 3.1 Prediction accuracy

In contrast to the traditional digital processors, photonic neural networks process mathematical models in analog way. Their numerical accuracy is sensitive to system flaws; therefore, the prediction accuracy of photonic neural networks will suffer from large system flaws. In this section, we show the impacts of two major system flaws (system noise and device imbalance) on the prediction accuracy of IPCNN. A simulator is built in a computer by the mathematical model of each component in IPCNN. A four-layer convolutional neural network model is chosen to conduct the MNIST-handwritten digit classification task. Figure 3(a) depicts the neural network model in detail. Two convolutional layers are cascaded by two fully-connected layers. The neural network parameters are trained in a digital computer, and the IPCNN simulator is adopted to execute convolutional layers in inference phase. The rest of the network is conducted by the digital computer.

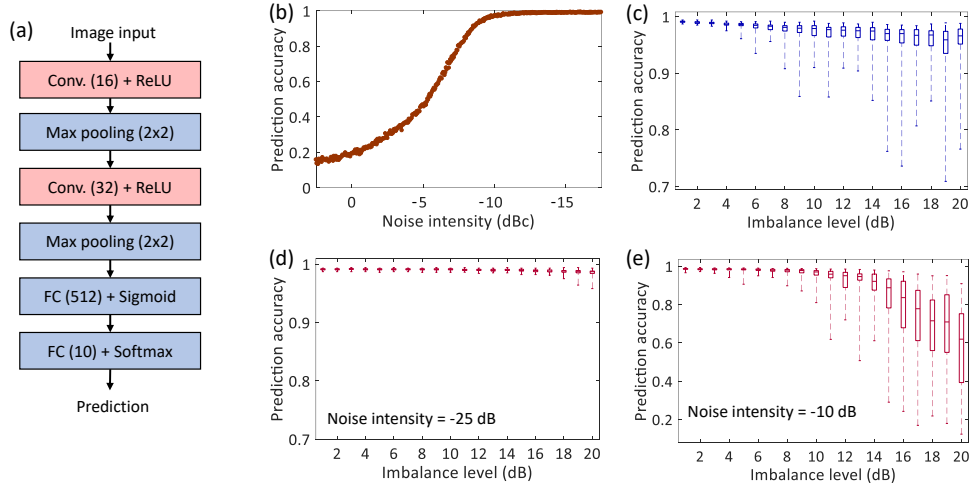

Fig. 3 Results of IPCNN prediction accuracy evaluations. (a) Neural network model used in the noise and imbalance evaluations. Linear operations and nonlinear activations are shown in each layer. "Conv. (32)" represents a convolutional layer with 32 output channels. "FC (512)" denotes a fully-connected layer with 512 output neurons. (b) Prediction accuracy dependence on noise. Noise optical intensity is normalized to the signal optical intensity. (c) Prediction accuracy dependence on imbalance level. Results are stochastic from 100-time simulations. (d) and (e) Prediction accuracy after digital calibration when the noise is -25 dBc and -10 dBc, respectively. Results are stochastic from 100-time simulations.

Firstly, we investigate the noise impact on prediction accuracy. The noise of IPCNN are majorly from lasers, photodetectors, TIAs, and the input electrical signals. Despite many different noise sources, the overall impact of these noises are given to most attention. Therefore, these noise sources are normalized to the relative noise-equivalent optical power (NEOP). For example, the NEOP of TIA can be calculated by the input-inferred noise current and the responsivity of PD. By setting the single optical power to be 1, the unit of NEOP is normalized to dBc. With this normalization, we can determine the prediction accuracy with an overall noise level. The results are shown in Fig. 3(b). With high NEOP levels, the prediction accuracy drops dramatically by the increasing of noise level. However, with NEOP levels lower than -10 dBc or signal-to-noise ratio (SNR) higher than 10 dB, the prediction accuracy stays over 97%.

Secondly, system imbalance is also a common problem of practical optical components. The imbalance would leave fixed errors in the computing results, degrading the prediction accuracy. In IPCNN, imbalances can be introduced by imbalanced laser output intensities, imbalanced insertion losses of modulators, imbalanced optical splitters, imbalanced photodetectors, etc. In the evaluation, overall imbalance level is considered, which describes the difference between largest output value and the lowest output value when the input data and kernels are always set to 1. Figure 3(c) gives the results of imbalance evaluations. Each box is the stochastic results of 100-time simulations at each imbalance level. We can find that with high imbalance levels, the prediction accuracy is quite uncertain. The probability of low accuracy gets larger with high imbalanced levels.

Note that the system imbalance leaves fixed errors to the results, the system can be calibrated with hardware tuning or digital processing to repair the imbalances. For those imbalances which cannot be hardware-tuned (e.g. splitting imbalance), it is necessary to add a digital process for the system calibration. If an output is larger or lower than desired value, it can be attenuated or amplified in digital data. In noiseless IPCNN, digital processing can perfectly restore the imbalances. However, in practical noisy systems, digital amplification would amplify noise as well. Therefore, we evaluate the impacts of noise and imbalance together to show the acceptable noise level and imbalance level in practical systems. Figure 3(d) and 3(e) shows the prediction accuracies when imbalances are digitally calibrated. Each box is the stochastic results of 100-time simulations at each imbalance level. At low noise level (-25 dBc), the digital calibration is useful. Large imbalances are well calibrated. At medium noise level (-10 dBc), large imbalances cannot be calibrated. In modern photonic integration technologies, it is reasonable to assume the overall imbalance level can be well controlled under several dB. Therefore, we can choose the noise level to be -10 dBc to maintain IPCNN working at a high prediction accuracy.

### 3.2 Scale and speed

In IPCNN architecture, the computing speed is majorly determined by hardware scale and modulation rate. The more splitting of optical power, the more computing can be done with a single input. However, the splitting scale is limited by the noise power, waveguide-capable power, and insertion loss. We evaluate the maximal scale ($C_OQ$) and computing speed in this section.

In IPCNN, the aggregate NEOP of photodetectors and TIAs are fixed. To maintain the SNR at 10 dB, large splitting scale requires large input optical power. However, there is a power bottleneck at the output of WDM and delay lines. The optical power conducted by a single waveguide is limited otherwise it will introduce nonlinear effect. We simulate a WDM and optical delay lines on $Si_3N_4$ platform. Shown in Fig. 4(a), the effective mode area of WDM output and delay line is 0.702 $\mu m^2$ and 1.599 $\mu m^2$, respectively. By considering the typical nonlinear index of $Si_3N_4$, $2.4\times10^{-19}$ $m^2/W$ [20], the effective nonlinear coefficient at 1550 nm is yielded as 2.77 rad/W/m and 1.21 rad/W/m for WDM output and delay line, respectively. Note that optical intensity distortion less than -12 dBc is acceptable to maintain a high prediction accuracy [17]. With the above nonlinear coefficients, optical power lower than 20 dBm is capable with the simulated structures.

To maintain the 10-dB SNR at photodetection, the NEOP of photodetection and the insertion loss from WDM output to photodetectors are critical. In Fig. 4(b), we inspect the relationship among maximal splitting scale, insertion loss, and NEOP of photodetection. Two points are marked on the image to show the maximal scale when NEOP is 6.3 $\mu$W. This NEOP is calculated given that the photodetector noise is 30 pW/$\sqrt{Hz}$ with 0.9 A/W responsivity [21] and the TIA input-inferred noise is 50 pA/$\sqrt{Hz}$ with 10 GHz bandwidth [22, 23]. With insertion loss less than 7.4 dB, the maximal scale supports IPCNN to conduct $C_O$=32, $Q$=9. Note that large insertion loss and NEOP could result in failure of IPCNN. The failure zone on the plot depicts the unacceptable NEOP and loss levels.

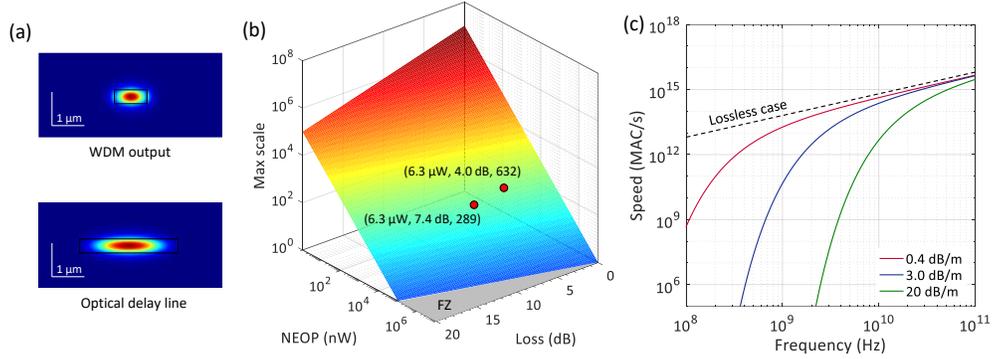

Fig. 4 Results of scale and speed evaluations. (a) Optical mode in WDM output and delay lines. The upper is a typical $Si_3N_4$ waveguide with size of $0.4\times1$ $\mu m^2$. The lower is widened waveguide with size of $0.4\times3$ $\mu m^2$. (b) Maximal splitting scale with different NEOP and loss. FZ, failure zone. (c) Computing speed with modulation rate. Three loss levels of delay lines are considered in the evaluation. The dashed line denotes the ideal speed when components are lossless.

From the maximal scale investigation, we evaluate the computing speed of IPCNN. We set the input channel number $C_I$=64 according to [24]. The $C_O$ and $Q$ parameters are determined by the scale evaluation results. Long delay lines introduce extra insertion loss and reduce the maximal scale, thus lowering the speed. With higher modulation rate, the computing speed increases because not only of the faster clock rate, but also of shorter required delay line. In Fig. 4(c), we show the relationship of computing speed and modulation rate at different delay line loss per unit [25, 26]. The insertion loss of splitters and MRRs is assumed to be 3.4 dB. With the increasing of modulation rate, the computing speed increases dramatically, approaching the lossless reference line. It is found that when $f_m$ is around 5 GHz, IPCNN can reach the speed of around 100 tera-MAC/s with low-loss delay lines.

### 3.3 Energy efficiency

Energy efficiency is one of the essential motivations of photonic neural networks, thus it is an important issue to perform the photonic advantage over digital electronic hardwares. In IPCNN, the adoption of optical delay lines mainly reduces the energy cost of E/O modulations. However, delay lines are another source of insertion loss, which degrades the achievable hardware scale. In this section, we show the energy efficiency evaluation results of IPCNN with typical integration technologies.

In IPCNN and other integrated photonic neural network architectures, energy costs are mainly from six parts: lasers, E/O modulations, thermal weighting matrix (phase shifter and MRR), photodetection (including TIAs), analog-to-digital conversions (ADCs), and digital processing circuits (e.g. the nonlinear operations). In this evaluation, we exclude the digital processing circuits since they are basically the same with all photonic neural network architectures. To evaluate the energy cost of each part, we estimate the energy budget from tail to head because it is necessary to maintain the 10-dB SNR at photodetection. The NEOP of photodetection is assumed to be 6.3 μW, the same as the above section and the corresponding power consumption of each TIA is 2.2 mW [22]. The insertion loss from the output of WDM to photodetection is estimated as 6.4 dB. With this insertion loss and NEOP, hardware scale of $C_OQ$=288 is feasible. Therefore, we set $C_I$=64, $C_O$=32, $Q$=9 in the evaluation. Given the reported power cost of each MRR, 19.5 mW [17], the total cost of VMMs is 359.4 W. Before the WDM output, E/O modulators and optical inputs also introduce insertion losses. We consider the modulators impose 4 dB loss and the $V_{pi}$ of them is 6V, which is typical for silicon-based Mach-Zehnder modulators. Since the modulator is supposed to work at full modulation depth, the power consumption of each E/O modulator is 90 mW in the 50-ohm system. The optical input port loss is considered as 2 dB. The total required optical power from lasers is 395 mW. Taking the typical wall-plug efficiency of WDM lasers, 5% [27-29], into account, the power consumption of lasers is 7.9 W. The power consumption of ADCs refers to an empirical review. That is 1 pJ/sample [30]. Table 1 lists the power consumption of all five parts and their corresponding proportions. Along with the energy evaluation of IPCNN, we evaluate the power consumption of other integrated photonic neural network architectures, including WDM-based ("DEAP" [17], "BW" [11]) and coherent-light-based ("Coherent" [10]). To maintain the same criteria, the power consumptions and insertion losses of components are assumed to be the same with the IPCNN evaluation. And the hardware scale, NEOP level, is also set at the same level with the IPCNN evaluation, i.e. $C_I$=64, $C_O$=32, $Q$=9, NEOP =6.3 µW. These architectures contain no delay line, thus the insertion loss of them is assumed 4 dB lower than IPCNN. Table 1 and table 2 also gives power consumptions of these reported architectures. Through assuming the modulation rate to be 5 GHz, we evaluate the energy efficiency of each architecture in Fig. 5. It is noticeable that all architectures perform worse than state-of-the-art electronic ASICs. This is because the thermal weighting matrices dissipate great portion of energy. If the weighting matrices are implemented with thermal components, the energy efficiency of photonic neural networks is unlikely to exceed electronics with current practical modulation rate.

Table 1. Power budget of components in IPCNN and DEAP

|  | IPCNN | | | DEAP | | |
|---|---|---|---|---|---|---|
|  | Power (W) | Ratio | Ratio (w/o weighting) | Power | Ratio | Ratio (w/o weighting) |
| Lasers | 7.96 | 2.1% | **54.8%** | 0.10 | 0.4% | 0.8% |
| E/O | 5.76 | 1.5% | 39.7% | 11.23 | **49.7%** | 98.9% |
| Weighting | 359.4 | **96.1%** | - | 11.23 | **49.7%** | - |
| TIA | 0.63 | 0.2% | 4.3% | 0.02 | 0.1% | 0.2% |
| ADC | 0.16 | 0.1% | 1.1% | 0.01 | 0.1% | 0.1% |

Table 2. Power budget of components in BW and Coherent

|  | BW | | | Coherent | | |
|---|---|---|---|---|---|---|
|  | Power (W) | Ratio | Ratio (w/o weighting) | Power | Ratio | Ratio (w/o weighting) |
| Lasers | 0.35 | 0.7% | 5.5% | 3.17 | 0.7% | 5.7% |
| E/O | 5.76 | 12.4% | **90.8%** | 51.8 | 12.5% | **93.8%** |
| Weighting | 39.93 | **86.3%** | - | 359.4 | **86.7%** | - |
| TIA | 0.07 | 0.2% | 1.1% | 0.07 | 0.0% | 0.1% |
| ADC | 0.16 | 0.3% | 2.5% | 0.16 | 0.1% | 0.3% |

Note that the weighting matrices of neural networks are static with no need for high-speed modulated. We should consider them as capacitive components (e.g. open-ended E/O MZM or MRR [31]) of zero power consumption at most time. This is also an original motivation of photonic neural networks [10]. Therefore, we evaluate the energy efficiency of integrated photonic neural network architectures with static capacitive weighting matrices. By eliminating the power consumption of weighting matrices, the energy efficiencies of photonic neural networks are promising to exceed the electronic state-of-the-art. In table 1 and table 2, we list the power proportions of each part with capacitive weighting matrices. In previously proposed architectures, the E/O modulation consumes most of the power. In IPCNN, optical delay lines reduce power consumption of E/O modulation by $Q$-times. Although the delay lines introduce insertion loss and the power budget of lasers is increased, the overall energy efficiency of

IPCNN can reach ~0.2 pJ/MAC, which is several times higher than previously proposed architectures.

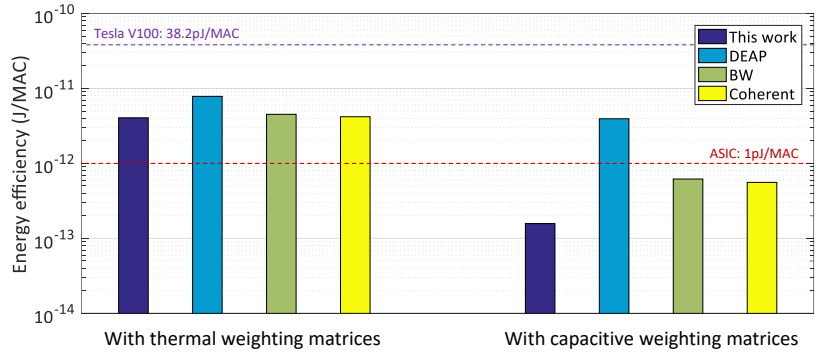

Fig. 5 Energy efficiency of different integrated photonic neural network architectures. We consider the weighting matrices to be thermal or capacitive. The red dashed line gives energy efficiency of electronic state-of-the-art [8, 9]. The purple dashed line gives the energy efficiency of a widely deployed commercial deep learning accelerator, Nvidia Tesla V100 [32].

## 4. Conclusions & Discussions

In this paper, we combine the technical advantages of optical delay lines and WDM to form an integrated photonic neural network architecture especially for convolutional layers. The electronic circuits are replaced with optical delay lines in order to conduct data manipulations. Consequently, the energy budget of input E/O modulation is reduced. With the application of WDM, the total length of required optical delay lines is dramatically reduced, thus becoming practically realizable with current integration technologies. To testify the feasibility and advantages of IPCNN, we evaluate its performance on prediction accuracy, the hardware scale, speed, and energy efficiency. Results shows that, with controllable insertion losses and noise level, IPCNN reach above 97% prediction accuracy on MNIST dataset. And it is potential to reach the speed of ~100 tera-MAC/s. In term of energy efficiency, IPCNN shows an obvious improve on E/O modulation. According to evaluation results, the overall energy efficiency is several times higher than previously proposed integrated photonic neural network architectures when the weighting matrices are implemented with capacitive components. Therefore, IPCNN is anticipated to be a potential method of high-energy-efficiency deep learning accelerator. We conduct the performance evaluations by mathematical and physical simulations in stochastic manner, with the purpose of validating the feasibility of IPCNN and providing an executable path of chip implementation from the system perspective. For the future, performance analyzes corresponding to each individual component are also important to bring IPCNN into real world.

From the system evaluation results, we conclude that there are two critical factors that mostly impact the performances of IPCNN. One is the noise level of photodetection. As a factor that determines the SNR of the system, lower noise level means a lower demand for optical power, a larger hardware scale, and higher energy efficiency. A potential improving method is to restrict the bandwidth of photodetectors and TIAs. For an IPCNN with modulation rate of $f_m$, a photodetection bandwidth of $f_m/2$ is sufficient. The restriction on bandwidth can improve the responsivity of photodetectors and lower the noise of TIAs. The second critical factor is insertion loss. Similar to the noise level, insertion loss dominates the hardware scale and SNR. IPCNN performances can be severely influenced by the insertion losses of laser inputs, modulators, and delay lines. To lower the insertion losses, it is achievable to adopt advanced integrated lithium niobate modulators [33, 34] and ultra-low-loss delay lines [35]. With the help of heterogeneous integration [36, 37] and low-loss assembly technologies, IPCNN can reach higher computing speed and energy efficiency. In addition to noise and insertion losses, IPCNN energy efficiency can be further enhanced by laser optimization, because the major portion of power dissipation is on lasers.

For now, a plethora of silicon-based photonic integration technologies supports the full fabrication of IPCNN on a single chip (except for lasers). In the future, photonic electronic hybrid integration technologies [38, 39] and the heterogeneous integration technologies are expected to provide opportunities for a full-functioning convolutional neural network. Moreover, for most integrated photonic neural networks, it is important that we consider the practical implementation and controlling methods [40] of ultra-large-scale weighting matrices. With the advantages of integrated optical circuits of high-parallelism and low-power, we are able to practically perform the computing speed and energy efficiency of IPCNN. Given that marginal computing terminals are a prevailing trend, this energy-efficient architecture is promising to benefit the high-performance deep learning inference on power-limited remote devices, including smartphones, drones, unmanned vehicles, and smart-cameras.

## 5. Acknowledgements

This work is supported by National Key R&D Program of China (Program No. 2019YFB2203700) and National Natural Science Foundation of China (Grant No. 61822508).

**Disclosures**

The authors declare no conflicts of interest.